# Optical pumping and relaxation of atomic population in assorted conditions

Swarupananda Pradhan[1,2,3] and Saptarshi Roy Chowdhury[1]

[1]Beam Technology Development Group, Bhabha Atomic Research Centre, Mumbai-400085, India

[2]Homi Bhabha National Institute, Anushaktinagar, Mumbai-400094, India

[3]Correspondence address: spradhan@barc.gov.in

**Abstract**

The precise control and knowledge over the atomic dynamics is central to the advancement of quantum technology. The different experimental conditions namely, atoms in a vacuum, an antirelaxation coated and a buffer gas filled atomic cell provides complementary platform for such investigations. The extent of changes in optical pumping, velocity changing collision and hyperfine changing collision rates associated with these conditions are discussed. There is a phenomenal change in the optical density by a factor of >25 times in presence of a control field in buffer gas environment. In contrary to earlier reports, we found confinement induced enhanced optical pumping as the mechanism behind the observed transparency in buffer gas cell. The feeble interplay of radiation trapping under specific conditions are pointed out. The diffusive velocity of atoms is measured to be ~25±12 m/s and ≤ 8±4 m/s for antirelaxation coated and buffer gas filled cell respectively. The studies will have useful application in measurements of relaxation rates, quantum memory, quantum repeaters and atomic devices.

## Introduction:

The working of quantum memory in warm atomic vapour rely on the associated elastic and inelastic processes involving the sample with its surrounding [1,2]. The relative values of these processes are a crucial parameter for realization of coherent matter waves [3]. The measurement of these parameters is generally carried out in magnetic and optical traps. The damping of forced spatial oscillation is a measure of elastic collision rate, whereas trap loss is related to inelastic rate. The Feshbach resonances are identified through enhancement in these rates as the magnetic field is varied [4]. The production of cold atomic sample and their investigation is an involved process. On the other hand, the warm atomic sample is a gateway to many areas of research like large-scale quantum entanglement, quantum memories, repeaters, quantum engineering and atomic devices [5]. The higher elastic collision rate is important for these non-isolated samples in preserving the spin coherence for a longer time. It directly influences the accuracy of quantum devices like atomic clock and atomic magnetometer [6, 7]. The various processes associated with laser atoms interaction and the decoherence mechanism plays pivotal role in most of the uses. The diffusive motion of atoms in these cells has gained continued-interest due to diverse applications [8]. We analyse the interplay of optical pumping with the relaxation rates in the steady state as well as temporal dynamics of atomic sample in different surroundings. The method for measurement of the associated rates and diffusive velocity are presented. We have used spatially separated pump and probe beam for measurement of diffusion velocity. This kind of geometry is used for highly sensitive magnetometer and the investigation will be useful for relevant research activities [9, 10].

The inelastic bouncing of the atoms from the wall is the limiting factor of coherent life time in an atomic cell. It is circumvented by two different approaches, one by coating the cell wall with an anti-relaxation material and the other by filling the cell with a buffer gas. The interaction between the atom and anti-relaxation material or buffer gas is mostly elastic in nature. The Rb atoms in uncoated atomic cell, anti-relaxation coated (ARC) cell and buffer gas filled cell provide three complementary paradigms for the study of atomic dynamics under the influence of resonant light fields. The Rb atom in uncoated cell is devoid of any other species in the cell. It undergoes feeble collision with other Rb atoms (low density) and gets bounced from the cell wall. The former is mostly elastic whereas the latter is highly inelastic in nature. The detrimental interaction with the cell wall is avoided by the ARC cell. The coating is expected to accompanied by small amount of residual gas due to chemical reaction between the Rb atoms and coating material. The amount of residual gas and its nature of collision with the Rb atoms is a matter of interest. On the other hand, the Rb atoms are confined inside the light beam for a prolonged time in the buffer gas filled cell. The high density of buffer gas leads to the pressure broadening. It is accompanied by a large frequency shift of the atomic resonances, a concern for metrological application. The uncoated cell and buffer gas filled cell can operate at high temperature. The stability of the coating material and its interaction with Rb atoms at high temperature is a shortcoming for the ARC cell. The study of atomic dynamics in these cell under identical condition is necessary for accessing their suitability in diverse area of research.

The experiment involves study of atomic dynamics by a probe beam tuned to the Rb D1 transition with a pump beam at Rb D2 transition. The investigation is divided into steady state behavior of the atomic spectrum and temporal decay of the optically pumped atoms. It carried out for pump beam in the probing volume as well as away from it and in absence of pump beam. The underlying observations allows us to visualize the different processes like optical pumping, velocity changing collision, hyperfine changing collision and the radiation trapping in these conditions. The temporal dynamics of ground state population is useful to extract the associated decay rates. The experimental conditions are changed to suitably extract these rates using a first principle model. The diffusive motion of the atoms is estimated using phenomenological observations. The methodology can open way for Feshbach spectroscopy in warm atomic sample.

## Experimental apparatus:

The experiments are carried out with Rb atoms in natural isotopic composition in (1) an uncoated cell prepared at $10^{-6}$ Torr residual gas pressure, (2) an ARC cell with octadecyltrichlorosilane coating with no intentional background gas, and (3) an uncoated cell with Ne atoms @ 50 Torr (Ne cell). All the cells have cylindrical shape with length ~50 mm and diameter ~25 mm. The experiments are carried out at room temperature ~$25^0$C.

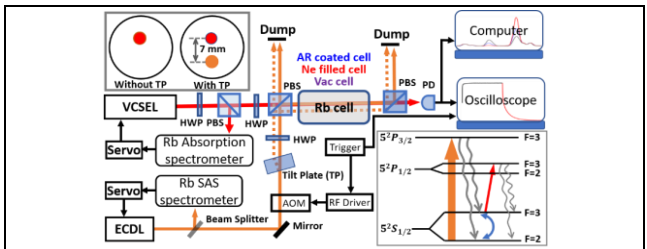

**Fig. 1:** Schematic diagram of the experimental set-up. The probe and pump beam are generated by the VCSEL and ECDL. The fast switching of the pump beam is done by an acousto-optic modulator (AOM). A tilt plate (TP) is used to spatially separate the pump beam from the probing volume as shown by the dashed orange line. **(top insert)**: The pump is in the probing volume in absence of TP. It is spatially separated by ~7mm in presence of TP. **(bottom insert)**: Relevant energy level of $^{85}$Rb atom. The pump beam (orange arrow) is locked to the D2 transition @780 nm. The probe beam (red arrow) scanned across or locked to the D1 transition @794 nm.

The energy levels of Rb-85 atoms with relevant optical coupling and decays are illustrated in the insert of Fig.-1. A Vertical Cavity Surface Emitting Laser (VCSEL) is tuned to the D1 transitions @794 nm ($5^2S_{1/2} \to 5^2P_{1/2}$) of Rb atom. A part of its beam (Fig.-1) is used for generating the frequency reference during the frequency scanning or for frequency stabilizing to the $^{85}$Rb $5^2S_{1/2}$, F=3 $\to$ $5^2P_{1/2}$, F'=3,2 transition, depending on the requirement. This $^{85}$Rb transition is written as F=3 $\to$ F' throughout the manuscript. Similarly, $^{85}$Rb $5^2S_{1/2}$, F=2 $\to$ $5^2P_{1/2}$, F'=3,2 transition is simply represented by F=2 $\to$ F' unless specified. The power of the remaining part of probe beam is controlled by a half wave plate (HWP) and a polarization beam splitter cube (PBS). The diameter of the probe beam is ~3 mm. The pump beam is derived from an external cavity diode laser (ECDL). It is locked to $^{85}$Rb $5^2S_{1/2}$, F=2$\to$$5^2P_{3/2}$, F'=3 D2 transition of $^{85}$Rb atoms using a saturation absorption spectroscopy (SAS) set-up. The first order diffracted beam by the AOM is used for the experiments. It is combined with the probe beam by a PBS and separated from the probe beam after passing through atomic cell by another PBS. The pump beam power is ~ 3mW with a beam diameter ~3 mm. For relevant experiments, a thick tilt plate (TP) in the path of the pump beam vertically separate its center from the center of the probing volume by ~7 mm. The probe transmission through the atomic cell is acquired as a function of laser frequency for the investigation of the steady state dynamics. Its temporal evolution with respect to the switching of the pump beam by an acousto-optic modulator (AOM) is acquired and averaged for ~10K sample by an oscilloscope. For this case, the probe frequency is locked to the F=3 $\to$ F' transition.

## Results and discussions:

*Steady-state behaviour:*

The spectrum of steady state optical density (OD) provides a picture of underlying physical process. The extent of these processes provides a guide line for the analysis of the temporal dynamics of the OD that is addressed at the later part of this article. The steady-state OD reflects the balance between the light induced population transfer and collisional thermalization. The light induced optical pumping is enabled by the pump beam, the probe beam and the scattered light through radiation trapping. These effects are discriminated in the study of temporal dynamics under specific condition. The collisional thermalization occurs through velocity changing and hyperfine changing process. These physical processes act to different extent depending on the utilized atomic cell. For the uncoated cell, the optical pumping by the probe beam can be neglected at smaller value of power due to limited transit time of the atoms. The steady-state OD spectrum for uncoated cell measured with probe power ~60 µW is shown in Fig-2. The spectrum for the pump beam off, in the probing volume, and at ~7 mm away from the probing volume provides a glimpse of the above processes. The measured OD is a function of isotopic abundance, population in the



ground level, absorption cross section, length of the cell, pump and probe intensity and their detuning.

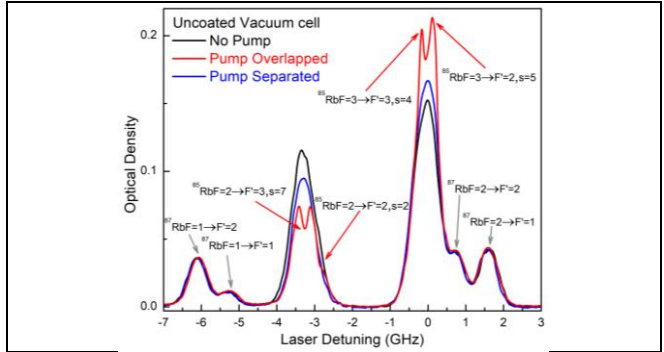

**Fig. 2:** The OD as a function of the probe laser detuning from the F=3 → F' transition for uncoated vacuum cell. The pump frequency is at $^{85}$Rb $5^2S_{1/2}$, F=2→$5^2P_{3/2}$, F'=3 transition. The probe beam power is ~60 µW.

The pump beam at $^{85}$Rb $5^2S_{1/2}$, F=2→$5^2P_{3/2}$, F'=3 transition has a line-width of ~1 MHz. It optically pumps a specific velocity group of atoms (near zero velocity) from F=2 to F=3 ground hyperfine level. Consequently, two Lamb dip associated with F=2 → F'=3 and F=2 → F'=2 transitions are seen for the pump located in the probing volume. The depleted atomic population appears as enhanced OD at F=3 → F'=3 and F=3 → F'=2 transitions. In earlier reports, observation of narrow resonances has been made for both pump and probe beam tuned to the D2 transition of Rb and K atoms [11]. The role of various processes including optical pumping in pump-probe spectroscopy has been discussed in several literatures [12]. However, unlike these prior arts, the sub-Doppler resonances are purely due to optical pumping in view of the utilized coupling scheme. The amplitude of the depletion or enhancement in the OD depends on the relative transition strength (indicated by "s" in Fig.-2) of the concerned coupling [13]. The width of the velocity selective resonances is a measure of the power broadened width. It shows ~57% of atomic population to be coupled with the light field. These group of atoms are completely transferred to the F=3 ground level. The diffusion of the fresh and bleached atoms to and out of the probing volume governs the steady-state population distributions. The diffusing atoms in to the probing volume, mostly comes after bouncing from the wall. The wall collision randomizes the velocity and hyperfine distribution to the equilibrium values. The OD remains at a non-zero value at the depleted position due to the influx of fresh (in population equilibrium) atoms to the probing volume. Only a fraction of the bleached atoms reaches the probing volume for pump beam placed away from it. Further, these optically pumped group of atoms undergoes collision with other Rb atoms prior to reaching the probing volume. Consequently, the sub-Doppler resonance are washed out in the profile of the OD. However, there is a significant change in the OD at F=3 → F'=3 and F=3 → F'=2 transitions compared to the corresponding equilibrium values in absence of the pump beam. The change in the value of OD due to the pump beam indicates that the equilibrium of population between the hyperfine levels has not reached, whereas wiping of the sharp resonances indicates equilibrium velocity distribution. The observations are consistent with higher elastic collision rate between Rb-Rb atoms compared to the rate of hyperfine changing collision. The uncoated vacuum cell is an ideal tool for sub-Doppler spectroscopy that rely on the tailored velocity group of atoms. It is normally carried-out with pump beam in the probing volume.

The inelastic interaction of the Rb atoms with the cell wall is significantly reduced by the OTS coating in the ARC cell. The atomic spin survives thousands of bouncing from the coated material [14]. Consequently, the atoms entering into the probing volume are no longer having equilibrium hyperfine population in contrast to the uncoated cell. Thus, the OD is close to zero for the F=2 → F' transition whereas it gets significantly enhanced for the F=3 → F' transition transition for overlapping pump and probe beam as shown in Fig.-3. The complete optical pumping of the atomic population from one hyperfine level to other has crucial role for realization of quantum memory [2].

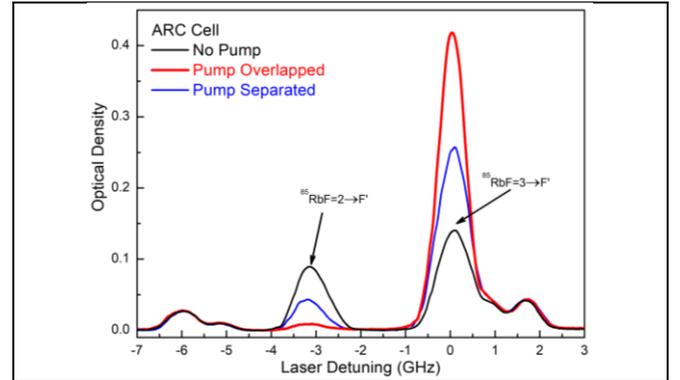

**Fig.-3:** The OD as a function of laser detuning for the ARC cell. The pump frequency is at $^{85}$Rb $5^2S_{1/2}$, F=2→$5^2P_{3/2}$, F'=3 transition. The probe beam power is ~60 µW.

The sharp sub-Doppler structures for the pump beam in the probing volume are not observed with the ARC cell, in contrast to the observation in the uncoated cell. The later also didn't show these sharp features for the pump beam spatially separated from the probing volume due to the velocity changing collision. It indicates rapid velocity changing collision in ARC cell compared to the uncoated cell. It can occur due to the additional gases in the atomic cell. The presence of such gases in paraffin coated ARC cell has been reported by following a different experimental procedure [15]. The reaction of Rb atoms with the OTS coating can also generates these background gases that leads to rapid equilibrium of the velocity distribution in the atomic sample. The change in OD for separated beams compared to overlapped beams is a measure of inelastic processes of the Rb atoms with coating and background gases. The achievement of velocity equilibrium with non-equilibrium hyperfine population distribution (for separated beams) revels stronger elastic collision of the Rb atoms with the surrounding



than the hyperfine changing collision. The ARC cell has unique uses for experiments that required longer state preservation time without introducing additional broadening. For example, the observation of multiple magnetic resonances due to a series of equally spaced light field in frequency domain can only be realized with ARC cell [16, 17]. The large homogeneous width in buffer gas filled cell plays detrimental role for observation of such resonances. In general, the ARC cell is useful for experiments involving quantum engineering, coherent population trapping and atomic devices. It has limited uses for sub-Doppler spectroscopy as the velocity selective resonances are rapidly washed out. An ideal ARC cell without any background gas can circumvent the shortcoming.

The buffer gas filled cell provides another paradigm where the interplay of radiation trapping has been intensely discussed. Radiation trapping infers to the scattering of a photon by an atom that is subsequently re-absorbed by another atom. It gives rise to a long range repulsive force that has been a subject on interest for the study of astrophysical objects to ultra-cold atoms [18,19]. In the context of buffer gas filled cell, radiation trapping has been assigned for the decrease in the optical pumping rate [20]. It stems from the impression of decrease in the absorption of resonant light as the atoms stays in the excited state for a longer average time. The process has been attributed to the decrease in Rubidium polarization with increase in atomic buffer gas pressure. We put forward a contradicting explanation for the miniscule OD in absence of pump beams in Fig.-4A. The OD is maximized in between F=2 → F' and F=3 → F' transition (see Fig.-4B) for probe beam power ~60 µW. The radiation trapping has limited explanation for such behavior. We attribute the confinement induced strong optical pumping by the probe beam for the observed decrease in OD. The optical pumping rate is much higher compared to the other two cells due to longer confinement of Rb atoms in the probing volume. The process is further augmented by larger homogeneous width of the resonances associated with the buffer gas filled cell. So even for a small power of probe beam (~60 µW), all the population are transferred to the uncoupled ground hyperfine level leading to decrease in its absorption. However, for the frequency of the probe beam in between the transitions, both hyperfine levels are coupled with the probe field and maximum OD is observed. We explicitly confirm this mechanism in the discussion after the following part.

For the overlapping beams, there is a minor reduction of OD at F=2 → F' as compared to the value in absence of pump beam. It indicates near complete optical pumping by the probe field itself. However, there is a large increase in OD at F=3 → F' in presence of the pump beam. A small increase in OD at $^{87}$Rb F=2→F'=1 transitions may be noticed in Fig.-4A. It arises due to the coupling of a fraction of $^{87}$Rb F=1 atoms with pump beam. The large homogeneous width associated with high pressure buffer gas is responsible for the observation. In contrast to the discussion involving ARC and uncoated cell,

the atoms coming to the probing volume are mostly from the diffusive motion from the neighbouring space and not from wall collision. The atoms also reside in the probing volume for a longer duration of time. Since the collision of Rb atoms with buffer gas is mostly elastic in nature, the sub-Doppler resonances are washed out along with a phenomenal increase in the OD at F=3 → F'. The presence of pump beam in the probing volume changes the OD by more than 25 times!! It can be further enhanced by increasing the operating temperature. This dramatic change is a special feature of confinement induced strong optical pumping associated with the buffer gas filled cell. It will have potential application in optical switch and optically controlled optical line filter [21, 22].

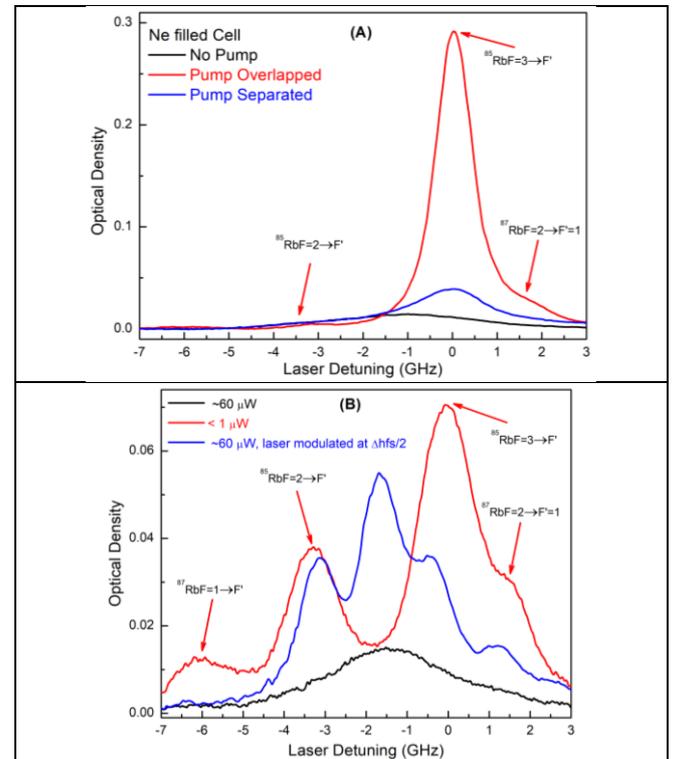

**Fig.-4:** The OD measured with Ne@50 Torr filled Rb atomic cell. **(A):** Strong optical pumping by the probe beam at ~60 µW leads to miniscule OD in absence of pump beam. The pump frequency is at $^{85}$Rb $5^2S_{1/2}$, F=2→$5^2P_{3/2}$, F'=3. **(B):** The OD is measured at probe beam ~60 µW (Black and blue curve) and <1 µW(red curve, average of 20 scan). The blue curve is obtained with probe laser frequency modulated at 1.518 GHz. The pump beam is off for these measurements.

The OD at F=3 → F' is reduced for separated beams due to combined action of optical pumping by the probe beam and hyperfine changing collision during diffusion of atoms from the pumping to the probing volume. This observation is similar to uncoated ARC cell, albeit with a larger change in the OD. The decrease in the OD due to optical pumping by the probe beam is further verified by reducing the probe beam power from 60 µW to sub-µW level as shown in Fig.-4(B).



The OD has substantially increased for low probe power but still much smaller compared to ARC and uncoated cell. It implies significant optical pumping even for very low probe power in buffer gas filled cell. The various transitions of the Rb atoms are partially resolved for sub-µW power of the probe beam. The $^{87}$Rb $5^2S_{1/2}$, F=1→$5^2P_{1/2}$, F'=1 and $5^2S_{1/2}$, F=2→$5^2P_{1/2}$, F'=2 transitions are not resolved due to larger homogeneous width in buffer gas filled cell.

The probe frequency is modulated at ~1.516 GHz to obtain a further insight in to the process. The chosen modulation frequency is half of the ground hyperfine splitting ($\Delta hfs$) for the $^{85}$Rb atoms. It creates few side modes separated by $\Delta hfs/2$ around the carrier frequency. During the scanning of the probe beam, for the carrier frequency tuned to one ground hyperfine level, the second neighbour side mode gets coupled with the other ground hyperfine level. The depletion in population by carrier frequency is relinquished by the side mode. Consequently, OD is increased even for probe power ~60 µW. There is a significant increase in the OD for the carrier frequency in between the F=2 → F' and F=3 → F' transition. Here, the neighboring side modes are simultaneously in resonance with both ground level leading to increased absorption of the laser beam. It suitably explains the optimal value of OD at the same location in absence of pump beam and modulation. The significant increase in OD for the pump beam at the probing volume or at lower probe power or for frequency modulated probe beam provides conclusive evidence of enhanced optical pumping in buffer gas filled cell.

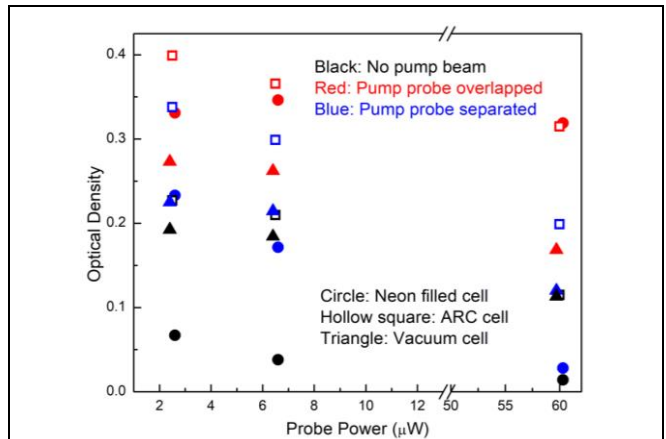

**Fig.-5:** The measured area under the profile of OD F=3 → F' for different probe beam power with pump frequency at $^{85}$Rb $5^2S_{1/2}$, F=2→$5^2P_{3/2}$, F'=3. The data are taken at probe beam power ~2.5µW, ~6.5 µW and ~60 µW at a constant pump power ~3 mW.

The widths of the resonances depend on the utilized cells owing to the diverse homogeneous width. The area under the spectral profile for F=3→F' transition (shown in Fig.-5) is an appropriate parameter for inter-comparison between the cells. The increasing OD with decreasing probe power is due to waning optical pumping by the probe beam. It is observed for all of the cell under three different condition of the pump beam. There is an exception for over lapped beam in buffer gas cell. The anomaly is verified by taking still smaller value of probe power (not shown) and requires further investigation to reach a suitable explanation. The change in all other data points is consistent with the earlier discussed physical processes. These mechanisms provide the basis for modelling the temporal dynamics of the OD after the pump beam is switched on or off.

*Temporal dynamics:*

The temporal dynamics of the OD is instrumental to extract the relaxation rates. The earlier studies are mostly limited to the temporal evolution of the optical rotation [23]. It is suitably explained by a double exponential decay model, having a faster and slower time scale. Here, we present an ab-initio model for the temporal evolution of the atomic population in the ground hyperfine level. The pump and probe beam are locked to the $^{85}$Rb $5^2S_{1/2}$, F=2→$5^2P_{1/2}$F'= 3 and F=3→ F' transition respectively. The fast switching of the pump beam is carried out with an AOM and the temporal evolution of the OD is monitored. The following rate equations describes the population dynamics.

$\frac{\partial n_1}{\partial t} = -\alpha_1 n_1 + \alpha_2 n_2 - \beta_1 n_1 + \beta_2 n_2 - \chi_1 n_1 + \chi_2 n_{11}$ (1)

$\frac{\partial n_2}{\partial t} = \alpha_1 n_1 - \alpha_2 n_2 + \beta_1 n_1 - \beta_2 n_2 - \chi_1 n_2 + \chi_2 n_{22}$ (2)

$\frac{\partial n_{11}}{\partial t} = -\beta_1 n_{11} + \beta_2 n_{22} + \chi_1 n_1 - \chi_2 n_{11}$ (3)

$\frac{\partial n_{22}}{\partial t} = \beta_1 n_{11} - \beta_2 n_{22} + \chi_1 n_2 - \chi_2 n_{22}$ (4)

$n_1$ and $n_2$ are the population of velocity group of atoms in F=3 and F=2 levels that are resonant with the probe and pump light field respectively. The experimentally measured OD represents $n_1$. Similarly, $n_{11}$ and $n_{22}$ are the velocity group of atoms in F=3 and F=2 levels that are not resonant with the light field. $\alpha_1$ and $\alpha_2$ are the optical pumping rate between $n_1$ and $n_2$. The population transfer between $n_1$ and $n_{11}$ ($n_2$ and $n_{22}$) occurs at a rate $\chi_1$ and $\chi_2$ through elastic collision. The population transfer between $n_1$ and $n_2$ ($n_{11}$ and $n_{22}$) occurs at a rate $\beta_1$ and $\beta_2$ through hyperfine changing collision. We may consider $\beta_2$=(7/5)× $\beta_1$ in view of the difference in the degeneracy number associated with F=3 and 2 level. Similarly, $\chi_2$=(4/3)× $\chi_1$ can be used in view of 57% of atomic population are observed to be coupled with the light field as measured for the uncoated cell. The set of nonlinear Eq. 1-4 will have many solution and it will be a difficult task to extract the associated parameters. However, analytical solution can be obtained for limiting experimental conditions.

It is convenient to visualize the dynamics of $n_1$ in concurrence with the above equations through three categories of physical processes as follows: (1)- A localized pumping to $n_1$ by the pump beam. It eventually changes the



global equilibrium of population. Although a specific velocity group of atoms are pumped, the high pumping rate ($\alpha_2$) and rapid velocity changing collision ($\chi_2$) ensures complete depletion of the lower coupled hyperfine level for limiting cases. (2)- The feeble optical pumping ($\alpha_1$) due to the probe beam. It leads to depletion of $n_1$ at the probing volume. (3)- Pure global actions like thermalization due to collision and feeble optical pumping due to the scattered pump light. The scattered light emanating from the pumping region counters to the optical pumping by the probe beam as it acts on the same near zero velocity group of atoms. It may be called radiation trapping and acts on the entire cell.

For vacuum cell, the short transit time of the atoms through the probing volume is the limiting factor. It provides a very short time window for investigation of the temporal dynamics. The atoms get randomized by colliding with the cell wall, once they escape the probing volume. Consequently, the signal amplitude is minuscule as shown in Fig.-6. The signal can be improved by increasing the cell temperature. However, it is expected to bring additional complexity due to higher Rb density and is beyond the scope of this article. We limit the comparison among the cells for identical condition only.

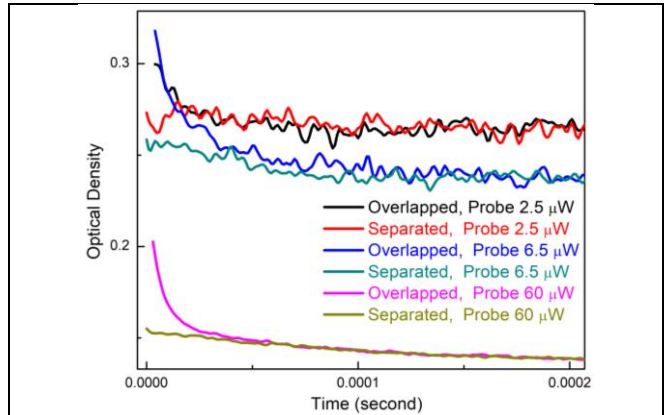

**Fig.-6:** The temporal evolution of the OD after the pump beam is switched off for different probe beam power. The data corresponds to uncoated atomic cell.

The simplest case is the probe beam at low power and spatially separated from the pump beam in the ARC cell. The pumped atoms get enough time to attain equilibrium Maxwellian distribution prior to reaching the probing volume as has been discussed in the steady-state dynamics. Thus, $\alpha_2=0$ and $\chi_1 n_1 = \chi_2 n_{11}$ can be taken in Eq.-1 for the evaluation of $n_1$ after the pump beam is switched off. It gives a simple analytic solution as

$$n_1(t) = \frac{1}{\alpha_1+\beta}\left[K_1 - A_1 \, e^{-(\alpha_1+\beta)t}\right] \quad (5)$$

where $K_1$ and $A_1$ are constant, and $\beta = \beta_1 + \beta_2$. The Eq.-5 assumes that $n_1 + n_2$ is constant and is appropriate for the atomic sample in equilibrium velocity distribution. At the probing volume, the steady-state is maintained by the continuous influx of optically pumped atoms against the depletion due to collisional thermalization. The optical pumping to $n_1$ stops after the pump beam is switched off. However, the influx of the pumped atoms continues even after the switching off due to the diffusive motion of the atoms. It allows to maintain a steady state for ~290±20 µs as shown in the Fig.-7A (insert). The diffusive motion is a measure of the background gas density in ARC cell. The pump and probe beams are separated by 7 mm, each having diameter of 3 mm. It gives separation between the pumping and probing region to be varying from 4 to 10 mm. The measured diffusive motion of the atoms is ~25±12 m/s. The accuracy of the measurement can be improved by using lower diameter of the beams. However, it will lead to compromise in the signal to noise ratio and will increase error in the measurement of transit time. The measured value of the diffusive velocity for ARC cell is much smaller than the rms velocity of ~240 m/s at the ambient temperature. The nearly constant value of $n_1$ for the initial time period indicates the negligible value of $\alpha_1$ for ARC cell at a probe power of ~2.5 µW. The curve for separated pump and probe beam fits well with Eq.-5 for data after 300 µs. The measured value of $\beta$ is ~10710 Hz with an assumption of $\alpha_1=0$. The $\beta_1$ and $\beta_2$ are expected to be related by $\beta_2=(7/5)\times \beta_1$.

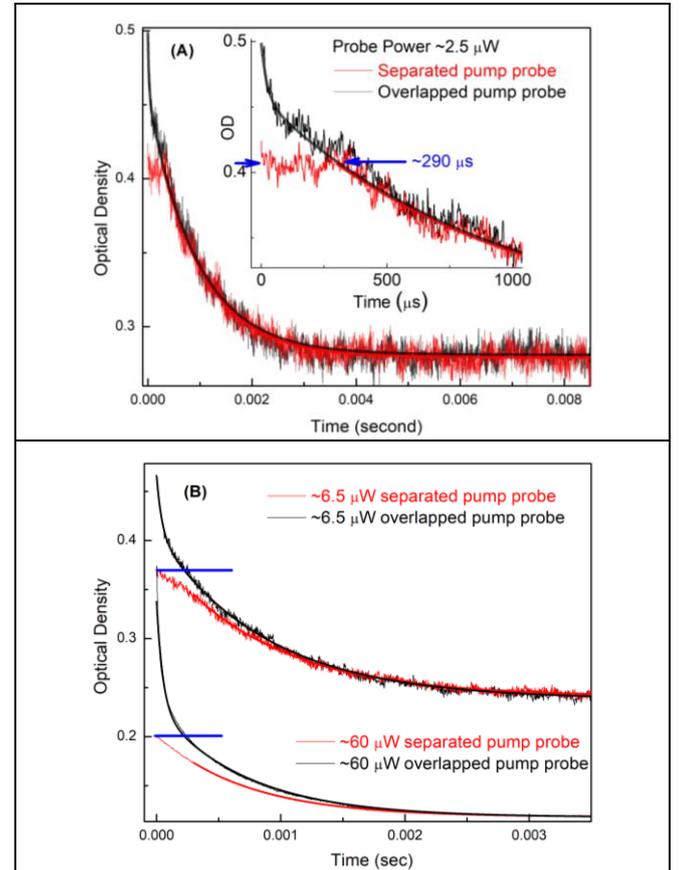

**Fig.-7:** The temporal evolution of the OD in ARC cell after the pump laser is switched off. The red and black thick lines are fitting to Eq.5 and 6. **(A):** The probe power is ~2.5 µW. The constant value of OD (red data) for the initial time period indicate negligible optical pumping by the probe beam. **(B):** The optical pumping by the probe beam is seen in the initial time of the red data set for probe beam power ~6.5 µW and ~60 µW.



For the pump beam in the probing volume, the initial dynamics is dominated by the velocity changing collision. Since the time scale for the velocity changing collision is much faster than $\alpha_1$ as well as hyperfine changing collisions ($\beta_1$ and $\beta_2$), the slower rates can be neglected for the initial dynamics. In other words, velocity changing collision and hyperfine changing process can be treated independent of each other as the former process comes to equilibrium very rapidly. Thus, the solution of Eq.-1 given by Eq.-5 can be extended to

$$n_1(t) = \frac{1}{\alpha_1+\beta}\left[K_1 - A_1\, e^{-(\alpha_1+\beta)t}\right] + \frac{1}{\chi}\left[K_2 - A_2\, e^{-\chi t}\right] \quad (6)$$

where $K_2$ and $A_2$ are constant, and $\chi = \chi_1 + \chi_2$. The Eq.-6 fits well with the data for overlapped pump-probe beam for probe power 2.5 µW. It gives the value of $\chi \sim 42608$ Hz by taking $\alpha_1=0$ and $\beta \sim 10710$ Hz. The measured $\chi \gg$ measured $\beta$ indicates the extent of validity of Eq.-6. The $\chi_1$ and $\chi_2$ are expected to be related by $\chi_2=(4/3)\times \chi_1$.

The curve for probe power 6.5 µW and 60 µW for ARC cell are shown in Fig.-7(B). The data for separated pump and probe beam fits well with the Eq.-5. The value of $\alpha_1 + \beta$ is 10785 Hz and 11030 Hz for probe power 6.5 µW and 60 µW respectively. However, the data for overlapped pump and probe beam starts to deviate from Eq.-6 as the value of $\alpha_1$ is increased. The fitting is good for 6.5 µW dataset, but significantly deviates for 60 µW data set as may be seen in Fig.-5B. The measured value of $\chi$ is 23400 Hz and 20200 Hz by fixing the corresponding values of $\alpha_1 + \beta$ for probe power 6.5 µW and 60 µW. These values of $\chi$ significantly differs from 42608 Hz obtained for probe power ~2.5 µW ($\alpha_1 \to 0$). Thus, for finite value of $\alpha_1$, the Eq.-6 is not valid and we cannot treat velocity changing collision process and hyperfine changing collision process in isolation. The dataset for 60 µW fits excellently with three exponential decay terms, but it is a challenging task to corelate with the relevant physical rates.

One of the important point to note in Fig.-7 is the similarity of the data at longer time scale for pump beam in and away from the probing volume. This resemblance shifts to longer time scale as the value of the probe power is increased. We have assigned the sharp change in the data to arrival of the final group atoms that has interacted with the pump light. The sharp deviation is smoothened due to optical pumping by the probe beam. The initial data point of the OD for separated beams would have remained unchanged till it met with the curve for overlapped beam in absence of optical pumping. It is shown by a horizontal blue line in Fig.-7B. The temporal position of the meeting of this blue line with the OD curve for overlapped beam can provide a rough estimate of diffusion time. We termed it as projection method. It may be noted that the diffusion time will be underestimated as optical pumping by the probe beam reduces the OD for overlapped beams. It occurs at ~245±10 µs for 6.5 µW and ~247±1 µs for 60 µW probe power for ARC cell. These values are comparable to the diffusion time ~290±20 µs measured for 2.5 µW ($\alpha_1 \to 0$). The projection method gives the lower limit of the transit time. The corresponding values of diffusion velocity are 30±12 m/s and 28 ±12 m/s. These values by the projection method are close to ~ 25±12 m/s measured in the limit $\alpha_1 \to 0$. The temporal dynamics for uncoated cell at 2.5 µW is indistinguishable from the noise level as in Fig.-6. We find diffusion time ~32±2 µs and ~22±2 µs for 6.5 µW and 60 µW probe power respectively, by applying the projection method. It gives a diffusion velocity 225±105 m/s and 330±170 m/s. Since the uncoated cell doesn't contain any additional gas, the diffusion time is expected to be same as the rms velocity (~240m/s). Thus, the projection method provides a good estimation of the diffusion velocity. The large error in the measurements involving uncoated cell compared to the ARC cell is due to shorter transit time.

The buffer gas filled cell is associated with large homogeneous width. The atoms are confined in the probing volume for a longer duration of time leading to enhanced optical pumping as has been seen in the steady-state behaviour. The strong optical pumping limits the uses of the Eq.-6 as has been discussed previously. The $\alpha_1$ is not negligible even for sub-µW pump power, as can be seen from the lower values of OD in Fig.-4B. Since the Ne density is much higher than the background gas in ARC cell, the diffusion time is also expected to be very long. Fig.-8 shows the temporal evolution of OD for Ne filled cell. The lower value of OD for probe power ~2.5 µW in Fig.-5 indicates the existence of strong optical pumping for buffer gas filled cell in absence of pump beam. Consequently, the sharp deviation of OD in the temporal dynamics is not observed in Fig.-8. Here, the projection method can be applied for estimating the lower limit of the diffusion time. The measured values are ~900±100 µs, 485±25 µs, 385±10 µs for 2.5 µW, 6.5 µW and 60 µW respectively. The corresponding diffusion velocity are 8±4 m/s, 15±7 m/s, and 18±8 m/s. The projection method gives the upper limit of diffusion velocity. Thus, diffusion velocity in atomic cell filled with Ne@50 T will be ≤ 8±4 m/s. The diffusion velocity for buffer filled gas is lower than ARC cell. It is consistent with the higher density of Ne gas compared to low pressure (expected) of background gas in the ARC cell.

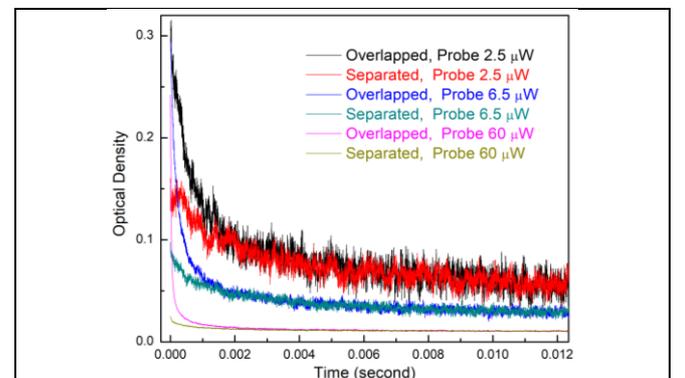

**Fig.-8:** The temporal evolution of the OD after the pump laser is switched off taken for Ne filled cell.

The experimental data for pump beam away from the probing volume follows Eq.-5 at longer time scale. The fitting is done with the data after 1.5 ms. The value of $\alpha_1 + \beta$ lies in the range ~310±50 Hz for probe power from 2.5 to 60 µW. It is much smaller then ~11000 Hz measured for ARC cell. The value of $\alpha_1$ for buffer gas filled cell is much larger than ARC cell. Thus, the in-elastic collision rate ($\beta$) for Rb atoms with Ne atoms is order of magnitude smaller than that with the background gas present in the ARC cell. The data with overlapped beam follows Eq.-6 for probe power of 2.5 µW. The measured value of $\chi$ is ~2000 Hz. This value is much smaller than ARC cell. It may be noted that buffer gas filled cell is associated with large homogeneous width of the transition. The measurement of $\chi$ using the decay curve relies on the discrimination of absorption for different velocity group of atoms. It breaks down for homogeneous width comparable or above the Doppler width. The Homogeneous width is larger than the Doppler width for the utilized Neon gas pressure of 50T. Thus, the measured value of $\chi$ doesn't reflect associated elastic collision rate and has little physical significance. For higher value of probe power, the data deviates from Eq.-6. It is similar to the observation for ARC cell as strong optical pumping limits the validity of underlying assumption.

The attributes of these cell have unique application in their merits. The uncoated cell is advantageous for study of hole burning, saturation absorption spectroscopy, velocity selective resonance and others. The ARC coated cell is more appropriate where longer state preservation time is required without introducing additional pressure broadening. The buffer gas filled cell is more suitable for longer state preservation time with a denser atomic sample as the cell can be heated to relatively higher temperature. The lower inelastic collision rate of Rb atoms with Ne atoms compared to the background gas in ARC cell will make it advantageous for relevant applications.

## Conclusions:

The optical pumping and relaxation rates of Rb atomic ensemble under complementary conditions are studied. The steady state behaviour provides a qualitative picture of the underlying mechanism whereas temporal dynamics allows measurement of these parameters. A suitable model is used and validity of the measurement for limiting conditions are discussed. We have explicitly shown the confinement induced enhanced optical pumping in buffer gas filled cell. It leads to optical control of OD by >25 times. The procedure for measurement of diffusion velocity is established. It can be directly measured for small optical pumping by the probe field. The projection method suitably estimates the upper limit of diffusion velocity for small optical pumping by the probe field. The diffusion velocity is found to be ≤ 225 ± 105 m/s, ~25 ± 12 m/s, and ≤8 ± 4 m/s for uncoated, anti-relaxation coated, and Ne@50T filled atomic cell respectively. The investigation will be useful in the widespread contemporary research area where atomic sample are manipulated by light field.

## Acknowledgements:

The authors are thankful to Shri. M. L. Mascarenhas and Dr. Archana Sharma for supporting the activity.

## References:

[1] R. Finkelstein, O. Lahad, I. Cohen, O. Davidson, S. Kiriati, E. Poem, and O. Firstenberg, Continuous protection of a collective state from inhomogeneous dephasing, Phys. Rev. X **11**, 011008 (2021).

2. D. Main, T.M. Hird, S. Gao, E. Oguz, D.J. Saunders, I.A. Walmsley, and P. M. Ledingham, Preparing narrow velocity distribution for quantum memories in room temperature alkali metal vapors, Phys. Rev. A **103**, 043105 (2021).

3. M. H. Anderson, J. R. Ensher, M. R. Matthews, C. E. Wieman, and E. A. Cornell, Observation of Bose-Einstein condensation in a dilute atomic vapor, Science **269**, 198 (1995).

4. J. L. Roberts, N. R. Claussen, S. L. Cornish, and C.E. Wieman, Magnetic field dependence of ultracold inelastic collisions near a Feshbach resonance, Phys. Rev. Lett. **85**, 728 (2000).

5. J. Kong, R. Jiménez-Martínez, C. Troullinou, V. G. Lucivero, G. Tóth and M.W. Mitchell, "Measurement-induced, spatially-extended entanglement in a hot, strongly-interacting atomic system", Nature Comm. **11**, 2415 (2020).

6. S. Pradhan, "Dual purpose atomic device for realizing atomic frequency standard and magnetic field measurement", US Patent No: 9097750 B2 (2015).

7. I.K. Kominis, T.W. Kornack, J.C. Allred, and M.V. Romalis, A sub-femto tesla multichannel atomic magnetometer, Nature **422**, 596 (2003).

8. R. Shaham, O. Katz, and O. Firstenberg, Quantum dynamics of collective spin states in a thermal gas, Phys. Rev. A **102**, 012822 (2020).

9. N. Wilson, P. Light, A. Luiten, and C. Perella, Ultrastable optical magnetometry, Phys. Rev. Applied **11**, 044034 (2019).

10. B. Pattons, E. Zhivun, D.C. Hovde, and B. Budker, All optical vector atomic magnetometer, Phys. Rev. Lett. **113**, 013001 (2014).

11. A. Banerjee and V. Natarajan, Saturated-absorption spectroscopy: eliminating crossover resonances by use of copropagating beams, Opt. Lett. **28**, 1912 (2003).

12. S. Pradhan and B. N. Jagatap, Magneto-assisted pump-probe spectroscopy of Cesium atoms, J. Opt. Soc. Am. B **28**, 398 (2011).

13. D.A. Steck, Rubidium 85 D Line Data, Rubidium 87 D Line Data, https://steck.us/alkalidata/.

14. S. J. Seltzer and M. V. Romalis, High-temperature alkali vapor cells with antirelaxation surface coating, J. App. Phys. **106**, 114905 (2009).

15. N. Sekiguchi and A. Hatakeyama, Non-negligible collisions of alkali atoms with background gas in buffer-gas-free cells coated with paraffin, Appl. Phys. B **122**, 81 (2016).

16. E. Breschi and A. Weis, Ground-state Hanle effect based on atomic alignment, Phys. Rev. A **86**, 053427 (2012).

17. R. Behera and S. Pradhan, Modulation induced splitting of the magnetic resonance, J. Phys. B: At. Mol. Opt. Phys. **53,** 205001 (2020).

18. T. G. Walker, D. W. Sesko, and C.E. Wieman, Radiation trapping force in magneto-optical trap, Phys. rev. Lett. **64**, 408 (1990).

19. S. Pradhan, Y. S. Mayya, and B. N. Jagatap, Velocity diffusion and radiation trapping force in a one-dimensional expansion of cold atomic clouds in a magneto-optical trap, Phys. Rev. A **76**, 033407 (2007).

20. M. A. Rosenberry, J. P. Reyes, D. Tupa, and T. J. Gay, Radiation trapping in Rubidium optical pumping at low buffer-gas pressures, Phys. Rev. A **75**, 023401 (2007).

21. S.L. Portalupi, M. Widmann, C. Nawrath, M. Jetter, P. Michler, J. Wrachtrup, and I. Gerhardt, Simultaneous faraday filtering of the Mollow triplet sidebands with the Cs-D1 clock transition, Nature Comm. **7**, 13632 (2016).

22. D. Pan, T. Shi, B. Luo, J. Chen, and H. Guo, Atomic optical stimulated amplifier with optical filtering of ultra-narrow bandwidth, Scientific Reports **8**, 6567 (2018)

23. M. T. Graf, D. F. Kimball, S. M. Rochester, K. kerner, C. Wong, D. Budker, E. B. Alexandrov, M. V. Balabas, and V. V. Yashchuk, Relaxation of atomic polarization in paraffin-coated cesium vapor cells, Phys. Rev. A **72**, 023401 (2005).